\documentclass[preprintnumbers,prd,twocolumn]{revtex4}
\usepackage{amssymb,amsmath}
\usepackage{graphicx,color}

\newcommand{\non}{\nonumber}
\newcommand{\half}{\tfrac{1}{2}}

\newcommand{\m}{ \mathcal{M} }

\begin{document}
\title{
A chiral random matrix model with 2+1 flavors at finite temperature and density
\footnote{Work supported in part by Grant-in-Aid of MEXT, Japan
(No.~19540269 and 19540273).}} 

\author{ H.~Fujii$^a$ and T.~Sano$^{a,b}$}
\address{%
$^a$Institute of Physics, University of Tokyo, Tokyo 153-8902, Japan\\ 
$^b$Department of Physics, University of Tokyo, Tokyo 113-0033, Japan 
}

\begin{abstract}
Phase diagram of a chiral random matrix model with the
degenerate ud quarks and the s quark at finite
temperature and density is presented.
The model exhibits a first-order transition at finite
temperature for three massless flavors,
owing to the U$_A$(1) breaking determinant term. 
We study the order of the transition 
with changing the quark masses and the quark chemical potential, 
and show that the first-order transition region expands as
the chemical potential increases.
We also discuss the behavior of the meson masses and 
the susceptibilities near the critical point. 
\end{abstract}

\date{\today}

\maketitle

{\it Introduction.---}
Study of the QCD critical point (CP)
\cite{Asakawa:1989bq,Barducci:1989wi,QCDCP}
is an intriguing fundamental issue
since its experimental confirmation will yield a strong
evidence for the QCD phase transition,
and energy-scan experiments searching for the QCD-CP
are being performed at Relativistic Heavy Ion Collider at BNL.
Although the existence of the critical point in the QCD phase diagram is
presumably inferred from the model studies and lattice QCD 
results\cite{Asakawa:1989bq,Barducci:1989wi,LQCD}, 
its absence is also a possibility\cite{deForcrand:2006pv,LQCD}.
In this paper, we adopt as a schematic model for QCD
the chiral random matrix (ChRM) model \cite{Sano:2009wd} 
which incorporates the U$_A$(1)-breaking determinant term\cite{KobaM,THooft}.
We report the phase diagram of this model
with the degenerate ud-quark mass $m_{\rm ud}$
and the s-quark mass $m_{\rm s}$
at finite temperature $T$ and quark chemical potential $\mu$.

The ChRM models have been successfully applied for qualitative study of
chiral properties of QCD\cite{ChRM,Halasz:1998qr}.   
In a ChRM model the Dirac operator on gluon field background is modeled 
by a matrix $D$ in the space of constant modes with small Dirac 
eigenvalues,
retaining the chiral symmetry $\{D,\gamma_5\}=0$.
The partition function of the model is given as an average of 
$\det{D}$ over random ensemble of matrix elements, 
which mimics the complexity of the
gluon dynamics. The finite $T$ and $\mu$ effects
are treated schematically as non-random 
external parameters appearing in $D$.
In Ref.~\cite{Halasz:1998qr},  the phase diagram of 
the ChRM model has been explored in the $T$-$\mu$ plane
and a tri-critical point (TCP) is found on the phase boundary
in the massless limit. 
The TCP changes to a simple CP when the quark mass 
is nonzero.
This result is consistent with the phase structure obtained in other model
studies with two quark flavors
\cite{Asakawa:1989bq,Barducci:1989wi,Berges:1998rc}
implying the scenario that the CP exists in the QCD phase
diagram as an endpoint of the first-order phase boundary.

Nature of the chiral transition in QCD is sensitive to the number of light
quark flavors, especially to the value of the s-quark mass $m_{\rm s}$.
Unfortunately, however, 
the phase structure of the conventional ChRM model\cite{Halasz:1998qr}
is independent of the number of flavors $N_f$.
In order to remedy this problem,
we have recently incorporated the U$_A$(1)-breaking determinant interaction
\cite{KobaM,THooft}
in the ChRM model \cite{Sano:2009wd}
by extending the zero-mode space \cite{JanikNZ97}
with the instanton gas model picture in mind.
This is the first ChRM model which describes 
the $N_f$ dependence of the chiral
transition allowing us to explore the phase structure 
varying the parameters $m_{\rm ud}$ and $m_{\rm s}$ 
in addition to $T$ and $\mu$.

{\it Model with determinant interaction.---} 
The chiral symmetry breaking manifests itself in the nonzero
density of the zero Dirac eigenvalues
through the Banks-Casher relation \cite{BanksC1979}.
The origin of small Dirac eigenvalues may be instanton configurations
of background gauge field and other nonperturbative gluon dynamics.
Here in our model we divide these fermionic modes into two categories,
$N_+$ and $N_-$ topological zero modes associated with 
$N_+$ instantons and $N_-$ anti-instantons respectively, and 
$2N$ near-zero modes generated by other complex 
dynamics \cite{JanikNZ97,Sano:2009wd}.
The Dirac operator $D$ then approximated with 
a martix of $2N + N_+ + N_-$ dimensions
with $N_+$, $N_-$ and $N$ being of the order of
the space-time volume ${\cal O} (V)$. 
The thermodynamic limit is taken as $2N + N_+ + N_- \to \infty$.
Note that $N_\pm$ should vary depending on the instanton distribution.

For fixed number of zero modes 
the model partition function is written in the chiral basis as
\begin{align}
Z_{N_+, N_-}=\int dR \; {\rm e}^{-N \Sigma^2 {\rm tr}RR^\dagger}
		 \prod_{f=1}^{N_f} \det(D+m_f) 
		 ,
\label{eq:fixedia}
\end{align}
with 
\begin{align}
D
=
\left(\begin{matrix}
     0       & {\rm i}R+C	\\
{\rm i}R^\dagger +C^T &   0
\end{matrix}\right), 
\end{align}
where $R \in {\Bbb C}^{(N+N_+)\times (N+N_-)}$ is a
random matrix following a Gaussian ensemble distribution with
the variance $1/(N\Sigma^2)$
and $C\in {\Bbb C}^{(N+N_+)\times (N+N_-)}$ 
is a matrix representing the effects of $T$ and $\mu$.
The matrix $D$ has $|N_+ - N_-|$ exact zero eigenvalues
when $R$ and $C$ are rectangular,
which is interpreted as a realization of the index theorem
in the ChRM model.
We adopt here the simplest form for $C$\cite{Halasz:1998qr}:
\begin{align}
C=
\left(\begin{matrix}
(\mu + {\rm i}T) \mathbf{1}_{N/2} & 0                         & 0\\
     0                     &(\mu - {\rm i}T)\mathbf{1}_{N/2}  & 0\\
     0                     & 0                        & 0
\end{matrix}\right)
\; ,
\label{eq:matter}
\end{align}
where $T$ and $\mu$ are schematic representation for 
the temperature and chemical potential effects, respectively.
Note that $D$ with $\mu \ne 0$ is non-Hermitian whereas
the partition function (\ref{eq:fixedia}) is 
still invariant under $\mu \leftrightarrow -\mu$. 
One should appreciate that 
the $N_+ \times N_-$ right-bottom block in $C$ 
corresponding to the topological zero modes
is set to zero.
This seems a reasonable assumption
if one notice that the finite $T$ and $\mu$ effects are introduced
as a boundary condition in the Matsubara formalism
and that the localized topological zero modes will be
insensitive to the boundary.
This discrimination is important indeed
in reproducing the physical $T$ dependence 
of the topological susceptibility\cite{Sano:2009wd,OhtaniLWH08}.

The complete partition function of the model
is obtained by summing over $N_+$ and $N_-$ with 
a distribution function $P(N_\pm)$:
\begin{align}
Z^{\rm RM}
=
\sum_{N_+, N_-} P(N_+)P(N_-) Z_{N_+, N_-}
\; .
\label{eq:total}
\end{align}
The $P(N_\pm)$ reflects a modeling of the instanton distribution
in the QCD ground state.
The authors of Ref.~\cite{JanikNZ97} adopted the
Poisson distribution, which involves arbitrarily large number
for $N_\pm$ and results in a model with no stable ground state.
Inspired by the lattice gas model within a finite box,
we instead  choose the binomial distribtion \cite{Sano:2009wd},
\begin{align}
P(N_\pm)=
\left(\begin{matrix}
\gamma N \\ 
N_{\pm}
\end{matrix}\right) 
\; 
p^{N_\pm} (1-p)^{\gamma N-N_\pm}
,
\label{eq:bino}
\end{align}
where $\gamma$ is a parameter of ${\cal O}(V^0)$ and 
$p$ is interpreted as the probability for a single instanton
to occupy a unit volume $V/(\gamma N)$. 
This distribution sets an upper bound $\gamma N$ for $N_\pm$
and gives rise to a stable effective potential
as a function of order parameters.
In fact, 
applying the standard bosonization procedure to (\ref{eq:total}), 
we find 
\begin{align}
Z^{\rm RM}
=&
\int dS \; {\rm e}^{- N\Sigma^2 \text{tr} S^\dagger S}
\nonumber \\ &
\times
\det{}^{\frac{N}{2}} 
\left [ (S+ \m)(S^\dagger+ \m^\dagger) - (\mu+{\rm i}T)^2 \right ]
\nonumber \\ &
\times
\det{}^{\frac{N}{2}} 
\left [ (S+ \m)(S^\dagger+ \m^\dagger) - (\mu-{\rm i}T)^2 \right ]
\nonumber \\&
\times 
\big [\alpha \det (S         + \m        ) +1 \big]^{\gamma N}
\big [\alpha \det (S^\dagger + \m^\dagger) +1 \big]^{\gamma N}
\nonumber \\
\equiv&
\int dS \;
{\rm e}^{-2N \Omega (S;\; T,\; \mu)}
\; ,
\label{eq:zrm}
\end{align}
where we defined the effective potential
$\Omega (S;\; T,\; \mu)$ in the last line.
$S \in {\Bbb C}^{N_f \times N_f}$ is
the order parameter matrix,
and $\m$ is the mass matrix.
The parameter $\alpha  = p/(1-p)$.
Note that the integrand of $Z^{\rm RM}$ is a polynomial of $S$
except for the exponential factor originating from 
the Gaussian ensemble distribution. Large values of
$S$ are suppressed by this Gaussian weight.

The determinant term with the coefficient $\alpha$ represents
the anomaly which  breaks explicitly the U$_A$(1) symmetry of the
effective potential $\Omega(S)$ even when $\m=0$.
For $S= \phi \mathbf{1}_{N_f}$ ($\phi \in {\Bbb R}$)
with $\m=0$, $\Omega$ simplifies to
\begin{align}
\Omega_{N_f} =& \frac{ N_f}{2} \left (
\Sigma^2 \phi^2 -
\half \ln [\phi^2 - (\mu+{\rm i}T)^2 ] [\phi^2 - (\mu-{\rm i}T)^2 ] 
\right ) 
\nonumber \\ & -
\gamma \ln |\alpha \phi^{N_f} +1|
\; .
\label{eq:pot3}
\end{align}
We see that the anomaly term yields $-\alpha\gamma \phi^{N_f}$
when expanded. In Ref.~\cite{Sano:2009wd} we studied this ChRM model
with two and three equal-mass flavors at finite $T$ 
with $\mu=0$, to show  a second- (first-) order phase transition
for $N_f = 2 (3)$.

{\it Phase diagram and meson masses with 2+1 flavors.---}
Choosing $S={\rm diag}(\phi_{\rm ud},\; \phi_{\rm ud},\; \phi_{\rm s})$
in the 2+1 flavor case with $\m={\rm diag}(m_{\rm ud},m_{\rm ud},m_{\rm s})$,
we have
\begin{align}
\Omega
&= 
\Sigma^2 \phi_{\rm ud}^2 -
\half (
\ln [\varphi_{\rm ud}^2 - (\mu+{\rm i}T)^2 ] 
+ (T\to -T) )
\non \\ &
+\half \left [
\Sigma^2 \phi_{\rm s}^2 -
\half (\ln [\varphi_{\rm s}^2 - (\mu+{\rm i}T)^2 ] 
+ (T \to -T) )
\right ] 
\nonumber \\ & -
\gamma \ln |\alpha \varphi_{\rm ud}^2\varphi_{\rm s} +1|
\; ,
\label{eq:pot21}
\end{align}
where $\varphi_{\rm ud}=\phi_{\rm ud}+ m_{\rm ud}$ and
$\varphi_{\rm s}=\phi_{\rm s}+ m_{\rm s}$.
The ground state is determined by the saddle-point condition
\begin{align}
\frac{\partial \Omega}{\partial \phi_{\rm ud}}
=0,
\qquad
\frac{\partial \Omega}{\partial \phi_{\rm s}}
=0
\; ,
\label{eq:gaps}
\end{align}
which becomes exact in the thermodynamic limit.

Prior to the numerical analysis,
we comment on the model parameters $\Sigma, \alpha$ and $\gamma$. 
Setting $\Sigma=1$ by redefinition of $S$ and other parameters, 
we searched such a set of parameters
$\alpha$, $\gamma$, $m_{\rm ud}$ and $m_{\rm s}$ 
that reproduces quantitatively the (ratios of the) meson masses
in the vacuum, but it was unsuccessful.
However, the model can describe the mass hierarchy
qualitatively as seen below in Fig.~\ref{f:masses}, and
we wish to study the model phase diagram 
as an schematic model for QCD. Note
that all the quantities are dimensionless in this work.

We also remark that the anomaly term makes a symmetry-broken
phase more stable. Indeed, 
no symmetry restoration occurs at finite $T$ 
for $\alpha \gamma > \Sigma^2 (=1)$ with $N_f=2$,
and the situation is similar even for $N_f=3$. 
Hence 
one must assume $\alpha\gamma \lesssim 1$ 
for study of the chiral restoration.

Finally, the transition at finite 
$\mu \ne 0$ with $T=0$ is first-order.
It is seen in the simple case (\ref{eq:pot3}) because
the symmetric phase $\phi=0$ and the broken phase $\phi > \mu$ 
are separated with the point $\phi=\mu$ where $\Omega = \infty$
or the integrand of $Z^{\rm RM}$ vanishes. This feature 
survives in more general cases with 2+1 flavors.

\begin{figure}[tbh]
\begin{center}
\includegraphics[width=0.47\textwidth]{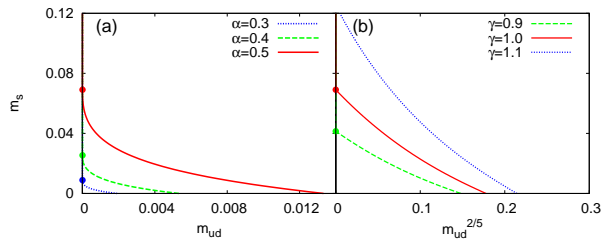}
 \caption{
The critical curves on the $m_{\rm ud}$-$m_{\rm s}$ plane 
(a) for $\alpha=0.3, 0.4, 0.5$ with $\gamma=1$ and
(b) for $\gamma=0.9, 1.0, 1.1$ with $\alpha=0.5$.
The finite-$T$ transition is first-order in the smaller-mass region
and  crossover in the larger-mass region. 
On the $m_{\rm s}$ axis with $m_{\rm ud}=0$, 
there is the $N_f=2$ chiral symmetry. 
The TCP is denoted by a dot for each parameter. 
}
\label{f:columbia}
\end{center}
\end{figure}

Let us study the phase diagram in 
the $T$-$m_{\rm ud}$-$m_{\rm s}$ space.
Since our model shows a first-order transition at finite $T$ 
for $m_{\rm ud}=m_{\rm s}=0$ and a crossover for large 
$m_{\rm ud}$ and $m_{\rm s}$\cite{Sano:2009wd}, 
there must be a line of a second-order transition
separating these two regions in the $m_{\rm ud}$-$m_{\rm s}$ plane.
This critical line is determined by the condition
$\Omega^{(n)}=0$ ($n=1, 2, 3$)
with 
$\Omega^{(n)} \equiv \partial^{n} \Omega/ \partial\phi_{\rm ud}^{n}$,
where $\Omega$ is a function of a single order parameter $\phi_{\rm ud}$
with $\phi_{\rm s}$ eliminated by the second equation in
(\ref{eq:gaps}). 
Note that
$\Omega^{(2)}=0$ is equivalent to 
$\det \partial^2 \Omega (S) / \partial \phi_i \partial \phi_j =0$
($i,j={\rm ud}, {\rm s}$), 
which implies the vanishing $\sigma$ mass (see below).

We present in Fig.~\ref{f:columbia} 
the critical line projected onto the $m_{\rm ud}$-$m_{\rm s}$ plane
for several values of $\alpha$ and $\gamma$.
When we increase the strength of the anomary term
$\alpha$ and/or $\gamma$, 
the region of the first-order transition expands.
For each parameter a TCP is found 
on the $m_{\rm s}$ axis,
where the $N_f=2$ chiral transition changes from a first to a
second-order one.
Near the TCP the critical line behaves as
$(m_{\rm s}^{\rm TCP} - m_{\rm s}) \propto m_{\rm ud}^{2/5} $ 
as is expected from the mean-field Landau-Ginzburg analysis,
which is clearly seen in Fig.~\ref{f:columbia} (b).
On the other hand, the line smoothly intersects the $m_{\rm ud}$ axis with
a finite slope$^{\rm \footnotemark[1]}$
\footnotetext[1]{We thank T.~Hatsuda's comment on this point.}. 
In fact, the model with the anomaly term
is symmetric under $m_{\rm ud} \leftrightarrow -m_{\rm ud}$
but asymmetric under $m_{\rm s} \leftrightarrow -m_{\rm s}$.

\begin{figure}[tbh]
\begin{center}
\includegraphics[width=7cm]{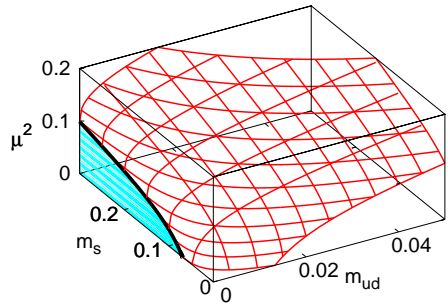}
 \caption{
Critical surface in the $m_{\rm ud}$-$m_{\rm s}$-$\mu^2$ space
with the parameters $\gamma=1$ and $\alpha=0.5$.
A series of TCP's is denoted by a thick line,
and the second-order transition with $N_f=2$ chiral symmetry
occurs in the shaded area.
}
\label{f:csurface}
\end{center}
\end{figure}

Next we extend our calculation to the finite $\mu$ case.
Fig.~\ref{f:csurface} exhibits the phase diagram in the 
$m_{\rm ud}$-$m_{\rm s}$-$\mu^2$ space.
We see that the region of the first-order transition expands
 as $\mu$ is increased.
This behavior indicates the existence of the CP in the
$T$-$\mu$ plane with the physical quark masses, 
provided that the finite-$T$ transition at $\mu=0$ is crossover. 
Varying the anomaly parameters $\alpha$ and $\gamma$,
we have confirmed that the expansion of the first-order region
with increasing $\mu$ is a robust result in our model
as far as we keep $\alpha$ and $\gamma$ constant.

The (screening) mass matrices for the scalar and pseudo-scalar mesons
are defined as the curvature of the potential around 
$\bar{S}={\rm diag}(\phi_{\rm ud},\; \phi_{\rm ud},\; \phi_{\rm s})$: 
\begin{align}
\left.
M_{ab}^{\rm s\; 2}
=
\frac{\partial^2 \Omega (S)}
	{\partial \sigma_a \partial \sigma_b}
\right|_{S=\bar S}
,
\quad
\left.
M_{ab}^{\rm ps\; 2}
=
\frac{\partial^2 \Omega (S)}
	{\partial \pi_a \partial \pi_b}
\right|_{S=\bar S}
\; ,
\label{eq:mass}
\end{align}
where $S=\lambda_a (\sigma_a + {\rm i}\pi_a)/\sqrt{2}$
with real parameters $\sigma_a$ and $\pi_a$,
and with $\lambda_a$ being the Gell-Mann matrices and 
$\lambda_0=\sqrt{\tfrac{2}{3}}{\rm diag}(1,1,1)$. 
See Appendix for explicit expressions for $M^2_{ab}$.
Because the quark mass term $m_{\rm ud} \ne m_{\rm s}$ breaks
the SU(3) flavor symmetry to cause 
nonzero mixing $M_{08}^2 = M_{80}^2 \neq 0$ in both 
the scalar and pseudo-scalar channels,
we diagonalize the matrices to get
the mass eigenvalues corresponding to $\sigma$ and $f_0$ for the scalars 
and $\eta$ and $\eta'$ for the pseudo-scalars.
One can show that the flat direction of 
$\Omega(\phi_{\rm ud},\phi_{\rm s})$ near the CP
coinsides with the $\sigma$ fluctuation direction.

In Fig.~\ref{f:masses}, we show the meson masses
as a function of $\mu$ at $T=T_c$
with parameters $\gamma=1$, $\alpha=0.5$,
$m_{\rm ud}=0.05$ and $m_{\rm s}=1.0$.
The model reproduces the empirical hierarchy of
the meson masses qualitatively in small $\mu$ region,
thanks to the anomaly term.
With increasing $T$ and/or $\mu$ the pseudo-scalar meson masses
remain nearly constant while the scalar ones, especially
the $\sigma$ mass, decrease.
At the CP $(T=T_c, \mu =\mu_c)$,
the $\sigma$ meson becomes massless. 
At higher $\mu>\mu_c$,
pairs of the masses $M_\sigma$-$M_\pi$, $M_\kappa$-$M_K$, 
and $M_\delta$-$M_{\eta'}$ get almost degenerate, 
reflecting the approximate $N_f=2$ chiral symmetry.

\begin{figure}[tbh]
\begin{center}
\includegraphics[width=7cm]{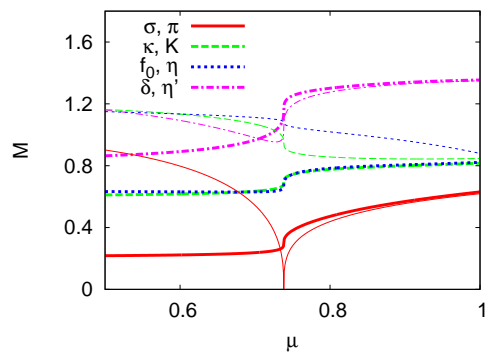}
 \caption{
Mesonic masses as functions of $\mu$ with $T=T_c$ fixed
for parameters $\gamma=1$, $\alpha=0.5$, $ m_{\rm ud}=0.05 $ and
$m_{\rm s}=1.0$.
The critical point locates at $(T_c, \mu_c)=(0.817, 0.738)$.
Thin (thick) lines indicate (pseudo-)scalar mesons.
}
\label{f:masses}
\end{center}
\end{figure}

Generalizing the mass matrix to
${\cal M}=\lambda_a(s_a+{\rm i} p_a)/\sqrt{2}$
with real sources $s_a$ and $p_a$, 
the scalar susceptibilities are defined as
\begin{align}
\chi_{ab}^{\rm s}
=
-\frac{\partial^2 \Omega(S({\cal M}); {\cal M})}
	{\partial s_a \partial s_b}
,
\label{e:sus_s}
\end{align}
where $S({\cal M})$ solves (\ref{eq:gaps}) for a fixed ${\cal M}$,
and the derivatives are evaluated at
${\cal M}={\rm diag}(m_{\rm ud},m_{\rm ud},m_{\rm s})$.
The pseudo-scalar susceptibilities
$\chi_{ab}^{\rm ps}$ are defined similarly as the response to $p_a$. 
We find the (pseudo-)scalar
susceptibilities $\chi_{ab}^{\rm s}$ and $\chi_{ab}^{\rm ps}$ 
in a diagonal form 
\begin{align}
\chi^{\rm s(ps)}_{ab}
=
\delta_{ab}\; 
\Sigma^2 \big(
\Sigma^2 /{M^{\rm s(ps)}_{aa}}^2 - 1 
\big)
\end{align}
for $a,b=1, \dots, 7$.
There is an mixing term for $a,b=0,8$ 
due to the SU(3) breaking:
\begin{align}
\chi^{\rm s(ps)}_{ab}
=
\Sigma^2 \left( 
\Sigma^2 ({M^{\rm s(ps)}}^2)^{-1}_{ab} - \delta_{ab}
\right) 
\; ,
\end{align}
which becomes diagonal when the mass matrix 
${M^{\rm s(ps)}}^2$ is transformed to be diagonal. 
Note that
the scalar susceptibility in the $\sigma$ channel diverges when the (screening)
mass $M_\sigma$ vanishes at the CP.
So do the quark number susceptibility 
$\chi_q=-\partial ^2 \Omega/\partial \mu^2$
as well as the `specific heat'
$\chi_T=-\partial ^2 \Omega/\partial T^2$,
through the mode mixing generated by the finite condensate
$\varphi_{\rm ud}$ and $\varphi_{\rm s}$ \cite{QTsuscept}.

{\it Conclusion.---}
As a schematic model for QCD, 
we have analyzed for the first time in the ChRM model 
the phase structure with 2+1 flavors 
at finite $T$ and $\mu$, 
which becomes possible by the inclusion of 
the U$_A$(1) breaking term\cite{Sano:2009wd}. 
We have drawn the critical curve separating 
the first-order transition region and the crossover region 
in the $m_{\rm ud}$-$m_{\rm s}$ plane with $\mu=0$,
and we have shown that 
the first-order transition region 
expands as the strength of the
anomaly term is increased.

Extending the model to the finite $\mu$ case,
we have shown the critical surface in the $m_{\rm ud}$-$m_{\rm s}$-$\mu$ space.
We have found that the first-order transition region expands
with increasing $\mu$, which is 
a supportive result for the existence of QCD-CP on the 
$T$-$\mu$ plane: 
one encounters a CP as $\mu$ is increased from zero,
provided that the finite-$T$ transition is crossover at $\mu=0$.
The meson mass hierarchy in the vacuum 
is qualitatively reproduced with the U$_A$(1) anomaly term and
the SU(3) flavor breaking. At the CP
the (screening) mass of $\sigma$ vanishes. 
Although we treat the model parameters independent of $T$ and $\mu$, 
possible rapid quenching of these parameters
at finite $\mu$ 
can give rise to a shrinkage of the first-order transition
region \cite{UA1inmedium}.
Furthermore, the role of the vector interaction between the quarks\cite{Vector}
deserves further study.  
These are beyond the scope of this schematic model.

{\it Appendix.---}
The mass matrix $M_{ab}^2$ is diagnal except for $a,b=0$ and 8.
For $a=1,2,3$, 
\begin{align}
M_{\delta,\pi}^2
	=&
	\Sigma^2
	\pm 
{\rm Re}\frac{\varphi_{\rm ud}^2 \pm z^2}{(\varphi_{\rm ud}^2-z^2)^2}
	\pm \gamma
	\frac{\alpha \varphi_{\rm s}}{\alpha \varphi_{\rm ud}^2 \varphi_s +1}
\; , 
\end{align}
and for $a=4, \dots, 7$,
\begin{align}
M_{\kappa, K}^2
	=&
\Sigma^2
\pm 
{\rm Re}
\frac{\varphi_{\rm ud}\varphi_{\rm s} \pm z^2}
     {(\varphi_{\rm ud}^2-z^2)(\varphi_{\rm s}^2-z^2)}
\pm \gamma
\frac{\alpha \varphi_{\rm ud}}{\alpha \varphi_{\rm ud}^2 \varphi_s +1}
\; , 
\label{e:kaon}
\end{align}
where $\varphi_{\rm ud}=\phi_{\rm ud}+m_{\rm ud}$, 
      $\varphi_{\rm s} =\phi_{\rm s} +m_{\rm s}$
and $z=\mu+{\rm i}T$. 
The upper (lower) sign corresponds to the (pseudo-) scalar meson.
The elements $M_{ab}^2$ for $a,b=0,8$ are written concisely in
another basis of $\lambda_{\rm ud}\equiv{\rm diag}(1,1,0)$ and
$\lambda_{\rm s}\equiv{\rm diag}(0,0,\sqrt{2})$, instead of $\lambda_{0,8}$:
\begin{align}
&M_{\rm ud,ud}^2 = \Sigma^2
	\pm 
{\rm Re}
	\frac{\varphi_{\rm ud}^2 \pm z^2}{(\varphi_{\rm ud}^2- z^2)^2}
\pm \gamma 
\frac{ \alpha^2 \varphi_{\rm ud}^2 \varphi_{\rm s}^2
      -\alpha \varphi_{\rm s}}
     { (\alpha \varphi_{\rm ud}^2 \varphi_{\rm s}+1)^2}
,
\\
&M_{\rm s,s}^2	= \Sigma^2
	\pm 
{\rm Re}
 \frac{\varphi_{\rm s}^2 \pm z^2}{(\varphi_{\rm s}^2 - z^2)^2}
\pm \gamma
        \frac{ \alpha^2 \varphi_{\rm ud}^4}
             {(\alpha \varphi_{\rm ud}^2 \varphi_{\rm s}+1)^2}
,
\\
&M_{\rm ud,s}^2 = M_{\rm s,ud}^2 
=
\mp \sqrt{2}\gamma \frac{\alpha \varphi_{\rm ud}}
                {(\alpha \varphi_{\rm ud}^2 \varphi_{\rm s}+1)^2}
\; .
\end{align}

\end{document}